\documentclass[aps,pra,amsmath,amssymb,twocolumn,groupedaddress]{revtex4-1}
\newcommand{\be}{\begin{equation}}
\newcommand{\ee}{\end{equation}}
\newcommand{\ba}{\begin{eqnarray}}
\newcommand{\ea}{\end{eqnarray}}
\newcommand{\baa}{\begin{eqnarray}}
\newcommand{\eaa}{\end{eqnarray}}
\newcommand{\ed}{\end{document}}
\newcommand{\lab}[1]{\label{#1}}
\newcommand{\re}[1]{(\ref{#1})}
\newcommand{\ci}[1]{\cite{#1}}
\def\pa{\partial}

\usepackage{graphicx}
\usepackage{color}

\usepackage{multirow}

\begin{document}

\title{Quantum dynamics of a hydrogen-like atom in a time-dependent box:\\ Non-adiabatic regime}

\author{S. Rakhmanov$^a$, D. Matrasulov$^b$, V.I. Matveev$^{b,c}$}

\address{$^b$ Physics Department, National Universty of Uzbekistan, Vuzgorodok, 100074,  Tashkent, Uzbekistan}
\address{$^b$ Turin Polytechnic University in Tashkent, 17 Niyazov
Str., 100095,  Tashkent, Uzbekistan}
\address{ $^b$ Federal Center for
Integrated Arctic Research, Ural Branch, Russian Academy of
Sciences, nab. Severnoi Dviny 23, Arkhangelsk, 163000
Russia}
\address{ $^c$ Northern (Arctic) Federal University, nab.
Severnoi Dviny 17, Arkhangelsk, 163002 Russia}


\vspace{10pt}

\begin{abstract}
We consider a hydrogen atom confined in time-dependent trap
created  by a spherical impenetrable box with time-dependent
radius. For such model we study the behavior of atomic electron
under the (non-adiabatic)dynamical confinement caused by the
rapidly moving wall of the box. The expectation values of the
total and kinetic energy, average force, pressure and coordinate
are analyzed as a function of time for linearly expanding,
contracting and harmonically breathing boxes. It is shown that
linearly extending box leads to de-excitation of the atom, while
the rapidly contracting box causes the creation of very high
pressure on the atom and transition of the atomic electron into
the unbound state. In harmonically breathing box diffusive
excitation of atomic electron may occur in analogy with that for
atom in a microwave field.
\end{abstract}

%
%
%
%
%
\maketitle

\section{Introduction}

Atoms and molecules confined in nanoscale domains have the
physical properties which are completely different than those of
free atoms. Such difference is caused by modification of the
boundary conditions imposed for quantum mechanical wave equations.
For free atoms the boundary conditions are imposed in whole space,
while for confined atoms one should solve the wave equations with
the boundary conditions imposed on a finite domain. Due to such
modification, properties of the atoms, molecules and matter depend
on the shape and size of a confining domain \ci{Sabin,Sil,Sen}.
Spatially confined atoms appear in many  physical systems, such as
optical traps \ci{Meystre,Gardiner}, low dimensional nanomaterials
\ci{Tsurumi}, fullerene \ci{Connerade1,Connerade2}, atom optic
billiards \ci{Raizen}-\ci{Douglas}, metallic hydrogen \ci{MH,MH1}
and  matter under high pressure \ci{Blank}. A key issue in the
study of such systems is the reaction of the electronic structure
of an atom (molecule) on the confining force \ci{Sil,Sen}. A
convenient model for the study of atoms under the spatial
confinement is the atom in  box system, where latter has spherical
shape and  impenetrable walls. Pioneering studies of atom-in-box
system  date back to the Refs. \ci{Michels,Sommerfeld}, where
effect of the pressure on an atom was explored in quantum
approach. Later, Wigner studied the problem  within the
Rayleigh-Schr\"odinger perturbation theory and showed that in the
limit of infinite box size, the result does not converge into that
for the free atom. Considerable number of papers on the
atom-in-box problem (see, e.g., \ci{Weil}-\ci{Burrows03} and
review paper \ci{Jaskolski} for more references) has been
published, since from these pioneering works. In \ci{Weil} the
problem of hyperfine splitting in such system is treated,
Ref.\ci{Rubinstein} presents first numerical solution of the
problem.  More comprehensive treatment of atom-in-box system can
be found in a series of papers by Burrows et.al
\ci{Burrows0,Burrows,Burrows03,Burrows04}, where the authors used
different analytical and numerical methods for finding eigenvalues
of the system. In \ci{Dragoslav,Lumb,Dragoslav1}  the quantum
dynamics of hydrogen atom confined in a spherical box and driven
by external electric field is studied.

Experimentally, atom-in-box system can be realized, e.g., in
co-called atom optic billiards which represent a rapidly scanning
and tightly focused laser beam creating a time-averaged
quasi-static potential for atoms  \ci{Raizen}-\ci{Montangero}. By
controlling the deflection angles of the laser beam, one can
create various box (billiard) shapes. Another method  is putting
the hydrogen atom inside the fullerene \ci{Connerade1,Connerade2}.
Different ways for confining of atom inside a cages in experiment
are discussed in \ci{Jaskolski}.

Usually, the studies of atom-in-box models  are mainly focused on
the static box boundaries. However, dynamical confinement appear
in different low-dimensional and atom-in-trap systems and the
response of the electronic structure of atom to the varying size
and shape of the ) confinement boundaries may play important role
in such systems. Some versions for experimental realization of
dynamical  confinement in different traps have already been
 considered in the Refs.\ci{TDT1,TDT2,TDT3}. In such cases the
dynamics of atomic electron is completely different than that for fixed boundaries.\\
In this paper we study quantum dynamics of one electron atom
confined in a spherical box with time dependent radius by focusing
on the response of atomic electron to the effect of moving walls
of the box. The time-dependence of the wall's position is
considered as non-adiabatic, i.e. we consider the cases of rapidly
shrinking, expanding and harmonically breathing boxes.

The problem of moving boundaries in quantum mechanics is treated
in terms of the  Schr\"{o}dinger equation with time-dependent
boundary conditions. Earlier, the quantum dynamics of a particle
confined in a time-dependent box was studied in different contexts
(see Ref. \ci{doescher}-\ci{Our02}). Here we consider similar
problem for an electron moving in a Coulomb field of the atomic
nucleus, confined in a spherical box with rapidly varying radius.
In such system central symmetry in the problem is not broken and
moving wall of the box plays the role external driving potential
for the hydrogen atom.

This paper is organized as follows. In the next section we give
brief description of the atom-in-box problem for a static box and
compare our numerical results with those by other authors. Section
3 presents detailed treatment of the quantum dynamics of the atom
confined in a spherical box with moving walls. Analysis of the
quantum pressure and force acting on the atom by moving wall of
the box is also presented. Finally, section 4 provides some
concluding remarks.
\begin{widetext}\label{fig:2}

{\bf Table 1.} The energy spectrum (first few $s-$states) of a
hydrogen atom confined in a spherical box with radius $r_0$ (in
atomic units).

\begin{table}[h!]
\begin{tabular}{|c|c|c|c|c|}
\hline \hspace{5mm}n\hspace{5mm} & \hspace{2mm}Confined atom
($r_0=10 $)\hspace{2mm} & \hspace{2mm}Confined atom ($r_0=20
$)\hspace{2mm} & \hspace{2mm}Confined atom ($r_0=100
$)\hspace{2mm} & \hspace{2mm}Free
atom\hspace{2mm} \\
\hline 1 & -0.4999992 & -0.4999995 & -0.5000000 & -0.5000000

\\
\hline 2 & -0.1128062 & -0.1249870 & -0.1249999 & -0.1250000

\\
\hline 3 & 0.0914223 & -0.0499180 & -0,0555555 & -0.0555555

\\
\hline 4 & 0.4051543 & 0.0167131 & -0.0312499 & -0.0312500

\\
\hline 5 & 0.8263889 & 0.1128777 & -0.0199999 & -0.0200000

\\
\hline 6 & 1.3511207 & 0.2372833 & -0.0138684 & -0.0138888

\\
\hline 7 & 1.9775747 & 0.3885954 & -0.0095963 & -0.0102040

\\
\hline

\end{tabular}
\end{table}

\end{widetext}

\section{The Hydrogen atom in a static spherical box}
Hydrogen atom confined in impenetrable spherical box was studied
in different papers by using different approaches for computing of
energy eigenvalues. In \ci{Capitelli,Cabrera} the eigenvalues
atom-in-box system are computed by solving the radial Schrodinger
equation numerically. Burrows and Cohen developed different
algebraic and approximate analytic approaches for finding
eigenvalues such system \ci{Burrows,Burrows04,Burrows}. In
\ci{Lumb} the authors used so-called  Bernstein-polynomials for
numerical solution of the radial  Schrodinger equation for atom in
spherical box. Review of the different approaches for finding of
energy eigenvalues of confined atoms can be found in
\ci{Jaskolski}.

Consider the hydrogen-like (one-electron) atom confined in a
spherical box with impenetrable walls with the radius $r_0$.
Assuming that the nucleus of the atom is fixed at the center of
box,
 for the dynamics of atomic electron in such system we have the stationary radial Schr\"{o}dinger equation which is given as (atomic units, $m_e =\hbar =e=1$ are used throughout this paper):
\be \Biggl[-\frac{1}{2}\frac{\partial^2 }{\partial
r^2}-\frac{1}{r}\frac{\partial }{\partial r}+\frac{l(l+1)}{2 r^2}
-\frac{Z}{r}\Biggl]R_{nl}(r) =E_{nl}R_{nl}(r), \lab{WE1} \ee where
$R_{nl}$ is the radial part of the wave function, $Z$ is the
charge of the nucleus, $n$ and $l$ are the principal and orbital
quantum numbers, respectively. The energy eigenvalues, $E_{nl}$
can be found from the  boundary condition for $R_{nl}(r)$
\ci{Michels,Dragoslav}:

 \be
R_{nl}(r)|_{r=r_0} =0. \lab{BC1} \ee

Unlike the unconfined atom, the energy spectrum of atom-in-box
system  is completely discrete. It should be noted that Eq.
\re{BC1} is not convenient for computing of the energy spectrum of
this system, as it does not allow to compute high number of
eigenvalues, especially, at small radii of the box. Here we
compute the energy levels of the atom-in-box system by
diagonalizing of the Hamiltonian (left part of Eq.\re{WE1})of the
system over the spherical box eigenfunctions basis. This method
allows to compute arbitrary high number of eigenvalues for any
value of the box radius.

Table 1 compares first seven energy levels for the atom in a
spherical box at different values of the box radius. Last column
in this table presents energy levels of unconfined (free) atom.
For the ground state level the eigenvalues of confined and free
atoms are the same, while for excited states they become
different. For confined atom the levels are  higher than those for
corresponding free atom energy levels.  Moreover, for very small
radius of the box the energy levels of atom-in-box system become
positive. Shrinking of the box size increases confinement, force
and pressure on the atomic electron, that causes the increase of
the average kinetic energy of the atomic electron leading to
positive eigenvalues in the spectrum of atom in box system.

Table 2 compares the energy levels obtained using our method with
those from the Ref.\ci{Burrows0}. As it can be seen, the agreement
is almost absolute (up to 11th order after the decimal place).

\begin{widetext}\label{fig:02}

\begin{table}[h!]
\centering {\bf Table 2.} The energy spectra  of a hydrogen atom
confined  in a spherical box computed in our approach  and
(underlined) results from \ci{Burrows0} at different values of the
box radius, $r_0$ (in atomic units).

\begin{tabular}{|c|c|c|c|}
\hline \hspace{1mm} States \hspace{1mm} &  \hspace{13mm} $r_0=8
$\hspace{13mm}
 & \hspace{13mm} $r_0=10 $\hspace{13mm}
&  \hspace{13mm} $r_0=14 $ \hspace{13mm}  \\
\hline
 \multirow{2}{*}{$1s$}  & -0.499 975 100 445 & -0.499 999 263 281 &
-0.499 999 999 498 \\ & \underline{-0.499 975 100 446} &
\underline{-0.499 999 263 282} & \underline{-0.499 999 999 498}

\\
\hline
 \multirow{2}{*}{$2s$}  & -0.084 738 721 356 & -0.112 806 210 295 &
-0.124 015 029 431 \\ & \underline{-0.084 738 721 357} &
\underline{-0.112 806 210 296} & \underline{-0.124 015 029 432}

\\
\hline
 \multirow{2}{*}{$2p$}  & -0.104 450 066 406 & -0.118 859 544 853 &
-0.124 540 597 990 \\ & \underline{-0.104 450 066 406} &
\underline{-0.118 859 544 854} & \underline{-0.124 540 597 990}
\\
\hline

\end{tabular}

\end{table}

\end{widetext}

\section{Hydrogen atom in a time-dependent spherical box}

Recent technological developments make possible trapping and
manipulating of atoms and molecules in time-dependent potentials.
Manipulation of the atomic Hamiltonians with both discrete and
continuum spectra is of practical importance in such field as
metrology and quantum information processing. Possibility for
creating of time-dependent traps and confining there particles and
atomic bound states have been discussed recently in different
contexts \ci{TDT1,TDT2,TDT3}.  Study of such systems requires
using effective models providing simplified and highly accurate
description of quantum dynamics and computing physically
observable quantities. One of such models can be atom confined in
a spherical box with time-dependent radius. Such time-dependence
does not break spherical symmetry of the system and one can
describe the whole system in terms of time-dependent radial wave
equation. Confining and cooling of atoms in time-dependent optical
traps was discussed in \ci{TDT1}. Creation of atomic Fock states
in time-dependent trap at different regimes of wall's motion was
studied in \ci{TDT2}. Ref.\ci{TDT3}presents proposal for confining
and driving  an ultrafast dynamics in a time-dependent box.
Quantum dynamics in externally manipulated time-dependent trap is
treated in \ci{TDT4}.

Here we consider atom confined in a spherical box with
time-varying radius given by $r_0 = r_0(t)$. In this case the
sphere retains its shape during the expansion(contraction), so
that the the central symmetry is not broken. Therefore, if atomic
nucleus is fixed at the center of sphere, the electron dynamics is
described by the time-dependent radial Schr\"{o}dinger equation
which is given as \be i\frac{\partial R(r,t)}{\partial
t}=\hat{H}R(r,t), \lab{TDSE1} \ee where \baa \hat{H}
=-\frac{1}{2}\frac{\partial^2 }{\partial
r^2}-\frac{1}{r}\frac{\partial }{\partial r}+\frac{l(l+1)}{2 r^2}
-\frac{Z}{r}.  \nonumber \eaa The boundary conditions for
Eq.\re{TDSE1} are imposed as
\baa R(r,t)|_{r=r_0(t)}=0. \nonumber
\eaa

To solve Eq.\re{TDSE1} one should  reduce  the boundary conditions
into time-independent form. This is can done by using the following
transformation \ci{mak91,mak92}: \be
 y=\frac{r}{r_0(t)}. \lab{tr1}
\ee

In terms of new coordinate, $y$ Eq.\re{TDSE1} can be rewritten as
\baa i\frac{\partial R(y,t)}{\partial t}=\Biggl[-\frac{1}{2
r_0^2}\frac{\partial^2 }{\partial y^2}-\Biggl( \frac{1}{2 r_0^2
y}-i\frac{\dot{r}_0}{r_0}y \Biggl)\frac{\partial }{\partial y}
\nonumber \\ +\frac{l(l+1)}{2r_0^2
y^2}-\frac{Z}{r_0y}\Biggl]R(y,t) \equiv \hat{\tilde H} R(y,t).
\lab{TDSE2} \eaa In Eq.\re{TDSE2} the self-adjointness is broken,
i.e., operator $\hat{\tilde H}$ is not Hermitian. In addition, due
to the first-order derivative in this equation makes complicated
its solution. Therefore, to restore Hermitticity and remove the
first order derivative, one can use the transformation of the wave
function which is given by \be
 R(y,t)=\frac{1}{r_0(t)^{3/2} y}e^{\frac{i}{2}r_0(t)\dot{r}_0(t)
 y^2}\Phi(y,t).
\lab{30} \ee  Doing such transformation  and  introducing of the
new time-variable defined as \ci{mak91,razavy,Our01}

\baa \tau =\int_0^{t}\frac{ds}{r_0(s)^2}, \nonumber \eaa

we reduce Eq.\re{TDSE2} into the Hermitian for which can be
written as \be i\frac{\partial \Phi}{\partial
\tau}=-\frac{1}{2}\frac{\partial^2 \Phi}{\partial y^2}+\Biggl(
\frac{1}{2}r_0^3 \ddot{r}_0 y^2+\frac{l(l+1)}{2y^2}-\frac{Zr_0}{y}
\Biggl)\Phi. \lab{TDSE3} \ee The boundary condition for $\Phi$ is
imposed as
\baa \Phi(y,t)|_{y=1} =0. \nonumber \eaa

We note that Eq.\re{TDSE3} can be obtained from Eq.\re{TDSE1}  by
using following unitary transformation for the Hamiltonian $H$
\ci{razavy}:

\baa \hat{\tilde H} =e^{-iV}e^{-iU}
(\hat{H}-i\frac{\partial}{\partial t})e^{iU}e^{iV}, \nonumber \eaa

where

\baa U=i(r\frac{\pa}{\pa r}+\frac{3}{2})\ln r_0(t), \nonumber \eaa

and

\baa V =-\frac{1}{2}r_0\frac{dr_0}{d\tau}y^2. \nonumber \eaa

\begin{figure}[t!]
\includegraphics[totalheight=0.21\textheight]{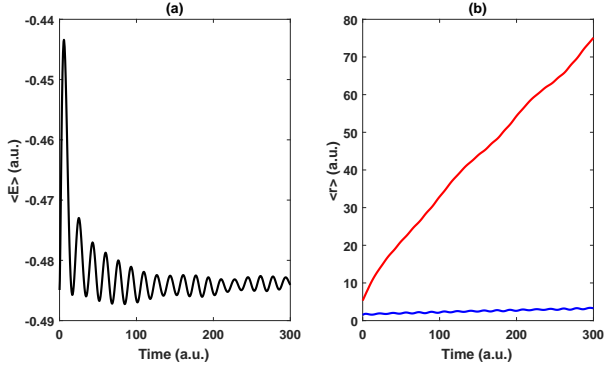}
 \caption{ (Color online) The average total energy (a) and coordinate (b) for hydrogen atom confined in a linearly expanding
 spherical box as a function of time for $a=10,\;b=1$. The average coordinate is plotted for two initial states, $1s$ (blue) and $2s$ (red).
 Atomic units are used in this and all other figures.} \label{fig:2}
\end{figure}

Eq.\re{TDSE3} is the Schr\"{o}dinger equation for an electron
moving in the field of Coulomb and time-dependent harmonic
oscillator potentials. The whole system is confined in a spherical
box with unit radius. Time and coordinate variables cannot be
separated in Eq.\re{TDSE3} and one needs to solve it numerically.
To do this we expand $R(y,t)$ in terms of the complete set of
eigenfunctions of a spherical box with unit radius:

\be \Phi(y,t) =\sum_{nl} C_{nl}(t)\varphi_{nl}(y),  \lab{tr22} \ee

where $\varphi_{nl}(y) =N_{nl}y j_l(\lambda_{nl}y)$ are the
eigenfunctions of the stationary Schr\"{o}dinger equation for a
spherical box  of unit radius, $j_l$ are the spherical Bessel
functions.  Inserting this expansion into Eq.\re{TDSE3} we get the
system of first order differential equations with respect to
$C_{nl}(t)$: \be i\dot {C}_{nl}(t) =r_0^{-2}\sum_{n'l'}
C_{n'l'}(t) V_{nln'l'}(t) +\varepsilon_{nl}r_0^{-2} C_{nl},
\lab{system1} \ee where $\varepsilon_{nl}$ are the eigenvalues of
the Schrodinger equation for spherical box and

\baa V_{nln'l'}(t) =<\varphi_{n'l'}|-\frac{Zr_0}{y}+\frac{1}{2}r_0^3
\ddot{r}_0 y^2|\varphi_{nl}>. \nonumber \eaa

In solving Eqs.\re{system1} numerically one should take into account
the normalization condition for the expansion coefficients,
$C_{nl}$:

\baa 4\pi\sum_{nl} |C_{nl}(t)|^2 =1, \nonumber \eaa

which follows from the normalization condition for the wave
function:

\baa \int_0^{r_0(t)}|\Psi(r,t)|^2 d^3r =1. \nonumber \eaa

\begin{figure}[t!]
\includegraphics[totalheight=0.21\textheight]{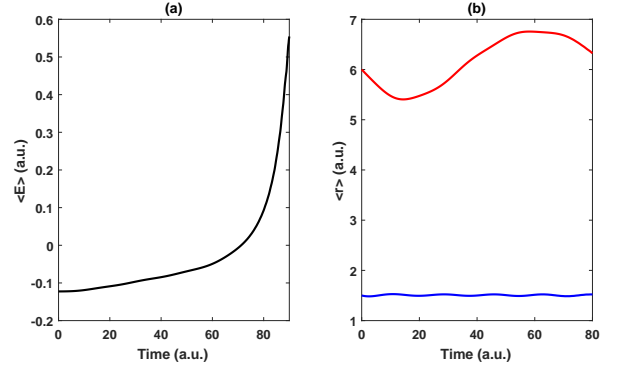}
 \caption{(Color online) The average total energy (a) and coordinate (b) for hydrogen atom confined in a linearly contracting
 spherical box as a function of time for $a=100,\;b=1$. The average coordinate is plotted for two initial states, $1s$ (blue) and $2s$ (red).} \label{fig:2}
\end{figure}

Having found the wave function, one can compute physical
characteristics of the atomic electron, such as average total
energy, force and pressure acting on atom by considering different
regimes of the wall's motion. In the following we will do that for
linearly expanding, contracting and harmonically oscillating box
wall.

Important physically observable characteristics of the atomic
electron is its average  total energy which is  defined as

\baa <E(t)> =4 \pi\int_0^{r_0(t)}R^*(r,t)
\Biggl(-\frac{1}{2}\frac{1}{r^2}\frac{\partial}{\partial
r}\biggl(r^2\frac{\partial}{\partial
r}\biggl)\nonumber \\
+\frac{l(l+1)}{2r^2} -\frac{Z}{r} \Biggl) R(r,t)r^2 dr,\nonumber
\eaa

and average kinetic energy given by

\baa <E_{kin}(t)> =4 \pi\int_0^{r_0(t)}R^*(r,t)
\Biggl(-\frac{1}{2}\frac{1}{r^2}\frac{\partial}{\partial
r}\biggl(r^2\frac{\partial}{\partial
r}\biggl)\nonumber \\
+\frac{l(l+1)}{2r^2} \Biggl) R(r,t)r^2 dr.\nonumber \lab{ekin}\eaa

In Fig. 1  the average total energy, $<E(t)>$ (a) and the average
coordinate (expectation value of the electron's position),
$<r(t)>$ (b) of the atomic electron are plotted for the hydrogen
atom confined in a linearly expanding spherical box ($r_0(t)
=a+bt$). At $t=0$ the box is very shrink ($a=10$), i.e. before the
expansion atom is highly compressed and kinetic energy of the
electron is very large. The average total energy in Fig. 1(a)
decreases in time, as box expands. This occurs due to the fact
that expansion of the box causes decrease in the kinetic energy of
electron. Decreasing is accompanied by oscillations. During long
enough time oscillation amplitude asymptotically goes to zero and
the value of $<E(t)>$ becomes close to that of the ground state
energy of (free) hydrogen atom. Such oscillations are caused by
the fact that at $t=0$, upon switching interaction between the
highly compressed atom and the moving wall of the box, the
transitions from the ground to excited states  accompanied by
mixing the states with different energies occur. As the box starts
to expand rapidly ($b=1$), the atomic electron, having very high
speed, starts to follow the moving wall. This can be seen from the
Fig.1 (b) , where the average coordinate of electron is plotted
for two initial conditions (the state of the atomic at $t=0$),
$1s$ and $2s$ states of the hydrogen atom in a spherical box. For
$2s$ state the growth is more rapid than that for $1s$.  Thus in
linearly expanding atom-in-box system where the expansion starts
from very compressed state, de-excitation caused by decrease of
the average kinetic energy of atomic electron occurs. When the box
expands up to very large radius, such de-excitation leads to the
transition of atom from highly excited to ground state. Therefore
putting (confining)of atom in an linearly expanding box can be
effective method for preparing ground state atoms.

\begin{figure}[t!]
\includegraphics[totalheight=0.20\textheight]{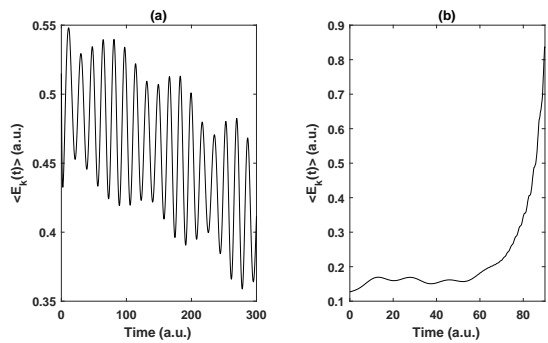}
\caption{  The average kinetic energy of atomic electron  for
hydrogen atom confined in a linearly expanding (a) and linearly
contracting (b) spherical box as a function of time for for the
parameters as those in Figs. 1 and 2, respectively.
 \label{fig:2}}
\end{figure}

Very important case  which is attractive from the viewpoint of
practical applications, is the contracting box. When box becomes
shrink, the pressure on the atom, created by the wall of the
shrinking box becomes higher. Such dynamic compression method
creating time-dependent pressure is used, e.g. in metallic
hydrogen formation experiments \ci{MH}. Dynamic high pressure
creates the extreme conditions for the atom when the force acting
by the wall on the atomic electron becomes higher than that of
atomic nucleus.

In Fig. 2  plots of the average total energy (a) and average
coordinate (b)  are plotted as  functions of time for linearly
contracting  ($r_0(t) =a - bt$) box with the initial radius
$a=100$ and contraction  velocity $b=1$. In computing of the
average energy, the initial state of atom is chosen as $2s$ state
of spherically confined atom. The average total energy grows
almost linearly during initial some stage, while upon shrinking up
to some (critical)size, the growth becomes abrupt. Further
shrinking leads to becoming the average total energy positive,
which is caused by growing of the average kinetic energy of the
atomic electron. Unlike the plot of Fig.1, there are no
oscillations in $<E(t)>$ for shrinking box. This is caused by the
fact that in this case, the "sudden switching" of the interaction
with the moving wall occurs at large distances from the atomic
electron and Coulomb center, which does not cause mixing of the
states. Plots of the average coordinate for two initial states,
$1s$ and $2s$ show that for $1s$ state the average coordinate does
not change, while for excited ($2s$) state $<r(t)>$ is not smooth
and the curve passes through the minimum values during the
shrinking.

The behavior of the average total energy presented in Figs. 1 and
2 can be understood, if one analysis the behavior of the average
kinetic energy as sa function of time. In Fig.3, time-dependence
of the average kinetic energy defined by Eq.\re{ekin} is plotted
for linearly expanding (a) and contracting (b) boxes
 at the same values of the parameters as in Figs. 1 and 2.
For linearly expanding box the average kinetic energy of the
atomic electron decays in time and the decay is accompanied by
oscillations. The curve of the average kinetic energy  for
linearly contracting box can be divided into two parts. In first
part $<E_{kin}(t)>$ manifests linear, but slow growth during some
initial time-interval, while in second part the growth becomes
abrupt. Such rapid growth can be explained by increasing pressure
on atomic electron by the wall of the box at small distances.

\begin{figure}[t!]
\includegraphics[totalheight=0.28\textheight]{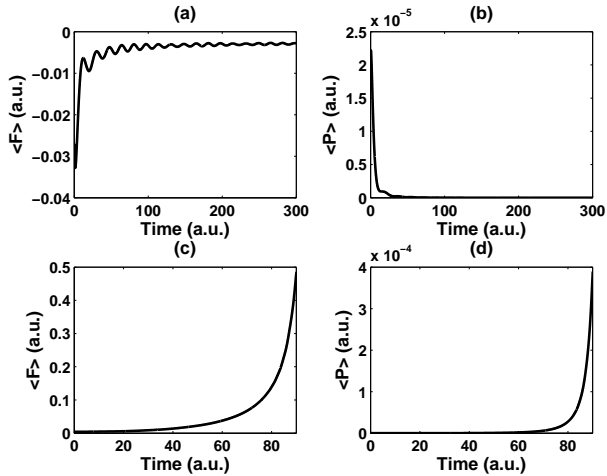}
 \caption{ Time-dependence of the average force (a), (c) and pressure (b), (d) acting on atomic electron by the wall of for  linearly expanding ($a=10,b=1$) and linearly contracting ($a=100,b=1$) spherical boxes.} \label{fig:2}
\end{figure}

An important characteristics  responsible for the behavior of
atomic electron in time-dependent box, are the average force
acting on the atomic electron by the moving wall and the pressure
caused by this force. The force operator for time-dependent box is
given by \ci{Our01,Our02} \be \hat F=-\frac{\partial\hat
H}{\partial r_0(t)}. \lab{for2}
 \ee
The average force can be calculated as \ci{Our01} \be
<F(t)>=-\frac{\partial}{\partial r_0(t)} \langle R(r,t)
|\partial\hat H | R(r,t) \rangle. \lab{for} \ee where $R(r,t)$ is
the wave function of the atomic electron in time-dependent box
determined by Eq.\re{TDSE1}.

Then the average pressure on the atom created by the moving wall can
be written as

\baa < P(t)> =\frac{< F(t) >}{4\pi r^2_0(t)}. \nonumber \eaa

Fig. 4 presents plots of the time-dependence of the average force
(a) acting on the atomic electron by the box wall and the average
pressure (b)  for linearly expanding box. Figs. 4c and 4d present
the similar plots for linearly contracting box. The values of the
parameters are the same as those in Figs. 1 and 2. For linearly
expanding box the modulus of the force acting on the electron by
moving wall decays on time, asymptotically approaching zero. The
value of the pressure for this system also decreases in time
approaching zero in long time limit. Such decrease causes the
decrease of the average kinetic energy of the atomic electron
which subsequently causes the behavior of the average total energy
observed in Fig. 1.  For linearly contracting box  the behaviors
of the average force and pressure are completely different than
those for expanding one. Sudden growth of the average force and
pressure starts  from some critical size of the box, $r_0(t) =25$.
This is very high pressure (few GPa) which can be achieved
dynamical compression method used, e.g. in metallic hydrogen
formation \ci{MH,MH1}. From the viewpoint of metallic hydrogen
physics, such pressure can cause appearing of "free electron gas"
before the formation of metallic hydrogen. Increasing of the
pressure for this case leads to the increase in the average
kinetic energy and to the behavior of $<E(t)>$ observed in Fig.3.

\begin{figure}[t!]
\includegraphics[totalheight=0.22\textheight]{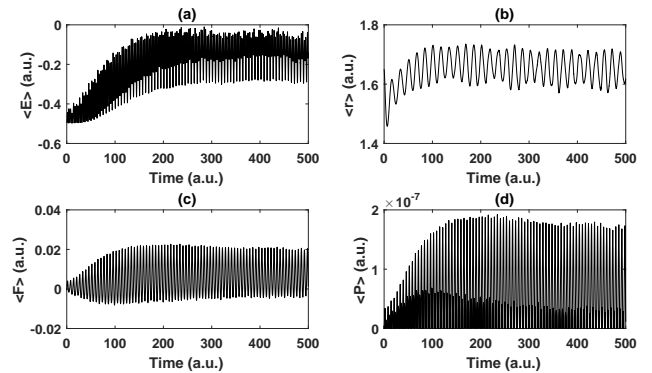}
\caption{  Time-dependence of the average total
 energy (a), coordinate (b), average force (c) and
pressure (d )acting on atomic electron by the wall for harmonically
breathing box
 ($a=100,b=10$,$\omega=1$) ) spherical box.}
\label{fig:4}
\end{figure}

\begin{figure}[t!]
\includegraphics[totalheight=0.22\textheight]{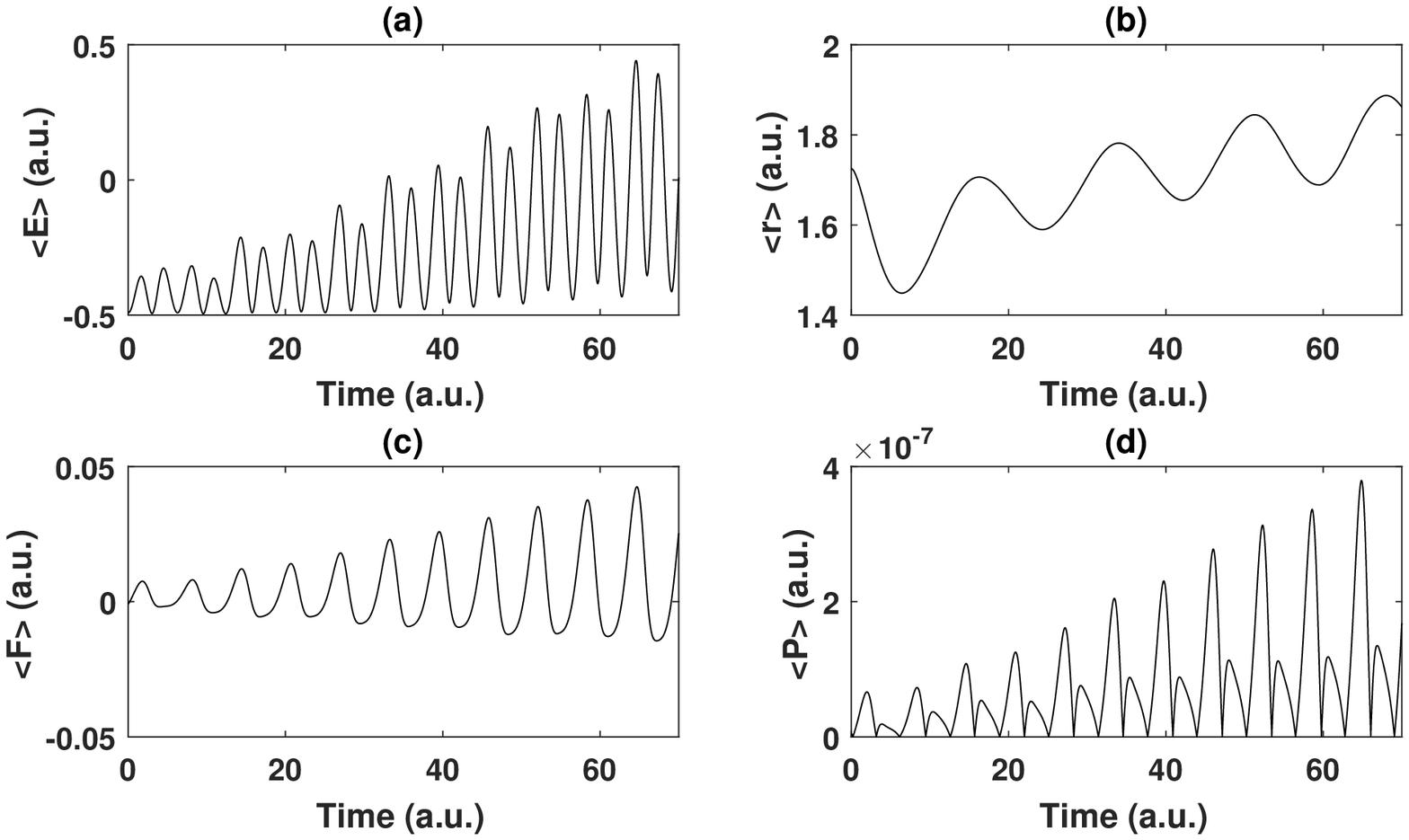}
\caption{ Time-dependence of the average total energy (a),
coordinate (b), average force (c) and pressure (d )acting on
atomic electron by the wall for harmonically breathing box
($a=100,b=15$,$\omega=1$) ) spherical box.} \label{fig:5}
\end{figure}

\begin{figure}[t!]
\includegraphics[totalheight=0.20\textheight]{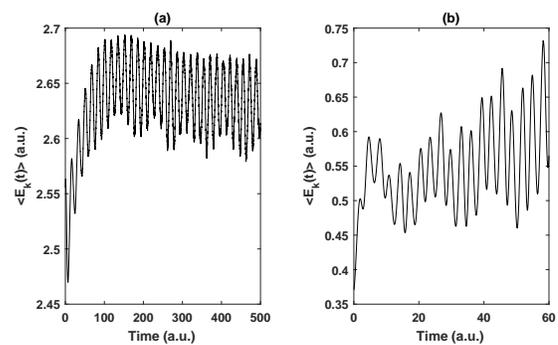}
 \caption{
The average kinetic energy of atomic electron  for hydrogen-like
atom confined in a harmonically oscillating box for the
oscillating parameters $a=100,b=10$,$\omega=1$ (a) and
$a=100,b=15$,$\omega=1$.} \label{fig:7}
\end{figure}

Explicit time dependence of $r_0(t)$ determines the nature of
confinement and is the key factor for the dynamics of atomic
electron. Therefore by choosing or tuning explicit form of
$r_0(t)$ one can achieve manifestation of different effects. The
above linear compression and expansion of the trap leads to
excitation and de-excitation of the atom. Another option for
tuning of the confinement dynamics is choosing of $r_0(t)$ as
time-periodic. In this case one deals with completely different
(than the above) behavior of the atomic electron. One of the
regimes belonging to this case, which is interesting from the
viewpoint of experimental realization (e.g., in atom optic
billiards), is harmonically breathing spherical box in where the
time-dependence of the box radius is given by $r_0(t) =a +
b\cos\omega t$, where $\omega$ and $b$ are the oscillating
frequency and amplitude, respectively. In Fig. 5 the expectation
values of the  total energy, coordinate force and pressure  are
plotted for $\omega =1$, $b=10$ and $a=100$. The average total
energy grows during some initial period after that suppression of
the growth can be observed and $<E(t)>$ does not become positive.
Similar "saturation" can be observed in the time-dependence of the
average force and pressure acting on atomic electron by moving
wall. The motion of the electron, characterized by the expectation
value of the coordinate, is localized around the value $1.7 a.u.$
However, as shows Fig. 6, for higher oscillation amplitudes the
interaction of atomic electron with the breathing wall leads to
diffusive excitation, i.e., the average total energy grows and
becomes positive after long enough time. Similar growth exhibit
the average force and pressure as a functions of time. Linear
growth (accompanied by oscillations)of the average coordinate can
be observed in this regime.  Thus there is a "critical" value
(approximately $12$a.u. in our case) of the breathing amplitude,
at which the average total energy changes its sign during some
time, It is clear that oscillation of the box's wall leads to
pumping of the energy into atom which causes its diffusive
excitation. In other words, atom in a box with oscillating walls
behaves itself as that in a monochromatic field, widely studied
earlier in the context of chaotic (diffusive) ionization
\ci{Casati,Jensen,Koch}.  Such model can be realized in atom
optics billiard where oscillating billiard boundaries can be
created by tightly focused laser beam with time-varying position
\ci{Raizen,Ariel2,Andersen,Rohwedder}.\\
Finally, to explain more clearly the behavior of the average total
energy  for harmonically breathing box, in Fig.7 we present the
plots of the average kinetic energy, $<E_{kin}(t)>$ as a function
of time.  For the set of parameters $\omega =1$,  $b=10$ and
$a=100$, the average kinetic energy (a)grows linearly during
initial period and the growth is suppressed after this period,
while for high amplitude (b) regime ($b=15$)  $<E_{kin}(t)>$ grows
linearly in time Such linear growth causes, of course, means
pumping of the energy into the atom that causes linear growth of
the average total energy (see, Fig. 6) and diffusive excitation of
atomic electron during long enough time interval. In the above
figures  characteristics of dynamic atom-in-box system as
functions of time. However, it is useful to explore dependence
these characteristics on the confinement radius, $r_0(t)$. For
linearly expanding and contracting boxes the curves do not change
their shapes, as such transformation imply linear deformation
only. However, for harmonically breathing regime the shape of the
curves are completely different, if one plots the dependence of
the averages as a function of $r_0(t)$. Fig.\re{curve1} presents
average total energy of hydrogen atom in harmonically breathing
spherical box as a function of $r_0(t)$ for to sets of parameters.
The curve consist of many cyclic lines whose origins and ends are
connected at the same point.

\begin{figure}[t!]
\includegraphics[totalheight=0.20\textheight]{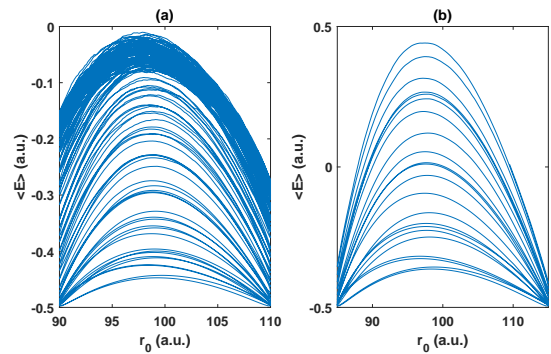}
 \caption{
The average total energy of atomic electron  for hydrogen-like
atom confined in a harmonically breathing spherical box as a
function of box radius, $r_0(t)$  for (a):  $a=100, b=10$,
$\omega=1$, $0 < t < 500$ and (b): $a=100, b=15$, $\omega=1$, $0 <
t < 500.$} \label{curve1}
\end{figure}

\section{Conclusions}
In this work we studied behavior of a hydrogen-like atom under the
dynamical confinement created by impenetrable spherical box with
time-dependent radius. The main focus of the study is given to the
response of the atomic electron's dynamics to the moving wall of
the box.

Time-dependence of the expectation values of the total energy,
kinetic energy, force, pressure and average coordinate are
analyzed for linearly expanding, contracting and harmonically
oscillating (breathing) boxes. For rapidly expanding box
de-excitation of the atomic electron leading to the decrease of
the average total energy occurs. Rapidly contracting box causes
creation of high pressure on atom and becoming the average total
energy positive. In case of harmonically breathing sphere, atom
behaves itself as that in microwave electric field, where the
diffusive (multi-step) excitation of atom occurs.

The above models can be used in  practically important problems,
such as metallic hydrogen formation, atom cooling and modeling the
behavior of matter under the extreme conditions caused by high
pressures. In addition, atom in spherical box with  rapidly
varying radius can be effective model  the study of different
ultrafast phenomena in atom optics, and atomic  attosecond physics
with trapped atoms. We note that the above approach is applicable
for the study of the adiabatic (slowly moving box walls) regime
which is the subject for the future project. Very attractive from
the practical viewpoint follow up of the above model is optical
harmonic generation in such system. If the trap boundary is
created by an optical field, interaction of such field with atomic
electron may cause harmonic generation of different orders. In
case of high harmonic generation, such a model with a dynamical
confinement could be very effective tool for attosecond pulse
generation.Moreover, earlier some results about the positive role
of the confinement in high harmonic generation have been reported
in the literature \ci{Dragoslav1,HHGA}.

\section*{Acknowledgement}
This work is partially supported by the grant of the Ministry of
Innovation Development of Uzbekistan (Ref. No.F-2-003).

\end{document}